\documentstyle[preprint,aps,amsfonts]{revtex}
\tighten
\draft
\widetext
\input epsf
\preprint{ITP-SB-99-03, CLNS 99/1610}
\bigskip
\bigskip

\newcommand{\beq}{\begin{equation}}
\newcommand{\eeq}{\end{equation}}
\newcommand{\beqs}{\begin{eqnarray}}
\newcommand{\eeqs}{\end{eqnarray}}
\newcommand{\lsim}{\mathrel{\raisebox{-.6ex}{$\stackrel{\textstyle<}{\sim}$}}}
\newcommand{\gsim}{\mathrel{\raisebox{-.6ex}{$\stackrel{\textstyle>}{\sim}$}}}

\begin{document}
\title{Collider Signatures from the Brane World}
\medskip
\author{Gary Shiu$^{1}$\footnote{E-mail:
shiu@insti.physics.sunysb.edu},
Robert Shrock$^{1}$\footnote{E-mail: shrock@insti.physics.sunysb.edu} 
and S.-H. Henry Tye$^{2}$\footnote{E-mail:
tye@mail.lns.cornell.edu}}
\bigskip
\address{$^1$Institute for Theoretical Physics, State University of
New York, Stony Brook, NY 11794\\
$^2$Newman Laboratory of Nuclear Studies, Cornell University, Ithaca, NY 14853}
\date{April 7, 1999}
\bigskip
\medskip
\maketitle

\begin{abstract}

{}We discuss some collider signatures of the brane world.  
In addition to the usual bulk (closed string) fields and brane 
(open string) fields in the Type I string picture, there are
closed string fields, namely, twisted modes, which are not 
confined on the branes but nonetheless are localized in the 
extra compactified dimensions.
While the coupling of the Standard Model (brane) fields with a
bulk mode (such as the graviton) is suppressed by powers of the 
Planck mass, their coupling to a twisted mode is suppressed only by 
powers of the string scale $M_s$, which can be as low as a few TeV.
This means these localized twisted fields can have important 
observable effects in the TeV range, including resonances in dijet 
invariant mass distributions in $\bar p p, pp \to$ jets + $X$. 
Given the current lower bound on the
fundamental higher-dimensional Planck scale, the experimental 
effects of these twisted fields may turn out to be larger than the 
effects of virtual and real KK gravity modes. The collider signatures 
of anomalous $U(1)$ gauge symmetries as well as other phenomenological 
implications of the brane world are also discussed. 

\end{abstract}
\pacs{}

\section{Introduction}

{}It was proposed in \cite{TeV} that the fundamental Planck scale 
can be around a TeV. In string theory, this implies that the string scale
$M_s$ is lowered all the way down to TeV scale\cite{lyk,anto,ST}. 
In this picture, the Standard Model (SM) fields reside inside of $p\leq 9$ 
spatial dimensional $p$-branes (or the intersection
of different sets of branes, as in \cite{ST}), 
while gravity lives in the higher (10 or 11)
dimensional bulk of spacetime. For $3<p<9$, this ``brane world'' scenario 
appears to be flexible enough so that various properties such as 
gauge and
gravitational coupling unification, dilaton stabilization, and the weakness of
the SM gauge couplings can be satisfied within
this framework \cite{BW}. The weakness of the four dimensional 
gravitational coupling is due to the presence of at least two large 
($\gg 1/M_s$) compact directions transverse to the $p$-branes on which the 
SM fields are localized. Issues such as coupling 
unification \cite{dien}, proton stability \cite{TeVphen}, 
neutrino masses \cite{neutrino}, 
fermion masses and mixing \cite{flavor}, 
and astrophysical/cosmological implications \cite{TeVphen,astro} 
have also been addressed \cite{related,zurab,quiros}. 
It has been argued that the
possibility of TeV scale gravity appears viable. 
If the fundamental Planck scale were around a TeV, 
clear signals due to strong gravitational interactions would appear at the 
LHC, extending existing constraints \cite{collider}. 
Furthermore, light Kaluza-Klein (KK) modes living in
the bulk can be produced. 

{}In general, the conventional Planck mass $M_P \sim 10^{19}$ GeV is 
related 
to the string scale by $(V_{p-3} V_{9-p} M_s^6)=(M_P/M_s)^2$, where 
$V_{p-3}$ and $V_{9-p}$ are the compactified volumes inside and 
transverse to the $p$-branes respectively.
Specifically, it is convenient to discuss the brane world picture in 
Type I string models, where brane fields are open string modes. 
The scenario given in Ref\cite{ST} is of this type. 
Here the Planck mass is given by
\begin{equation}
M_P \sim  M_s (M_s R_1) (M_s R_2) (M_s R_3) \\
        \sim M_{Pl} (M_{Pl} R_1)
\end{equation} 
where $R_i$ are the three radii characterizing the 3 compactified tori and 
$M_{Pl}$ is the 6-dimensional Planck mass scale \cite{TeV}, 
in the case with $n=2$ large dimensions. To stabilize the dilaton 
expectation value (and maybe even to induce SUSY breaking), the 
string coupling $g_s$ is likely to be strong (or, more
precisely, 
$g_s {\ \lower-1.2pt\vbox{\hbox{\rlap{$>$}\lower5pt\vbox{\hbox{$\sim$}}}}\ } 1$, and the theory appears not to have a dual weak coupling description).
For $p=5$, a typical squared gauge coupling $\alpha=g^2/(4 \pi)$ is given
by $\alpha \sim g_s/(M_s R_i)^2$ for $i=$2 or 3.  
To obtain the weak SM gauge couplings from 
a generic strong string coupling
requires that 
$(M_s R_2) (M_s R_3) \sim 10-100$.  
For $n=2$, a constraint from supernova
cooling requires $M_{Pl} > 30-50$ 
TeV \cite{TeVphen,SN1987A}, implying that $M_s > 3-20$ TeV. 
This value for $M_s$ is compatible with the SUSY 
models \cite{zurab} and the brane world picture \cite{BW}.
In this case, neither the Tevatron nor LHC energies may 
be high enough to detect 
either the strong gravity effect or the gravity KK modes. 
Are there other signatures at these energies that may be used to test the
brane world scenario?  The answer is yes, as we show here. 

{}The collider signatures that we discuss here 
do not arise from strong gravity or bulk KK modes (as in 
Refs. \cite{collider,nath}) but come from
two generic features of Type I string theory, which should also
be generic in 
the brane world picture: \\
\noindent $\bullet$ 
(1) In contrast to perturbative heterotic string theory, in some
Type I string models (which can be realized as Type II orientifolds and
are dual to non-perturbative heterotic strings), 
there can be more than one anomalous $U(1)$ gauge fields. 
They must be massive, though some of them can 
be relatively light and may be produced in high energy colliders 
as $Z^{\prime}$ vector bosons. \\ 
\noindent $\bullet$
(2) There are closed string modes which are not bulk modes, that is, 
they do not propagate freely in the bulk. Some of these modes are 
localized 
away from the branes so their couplings to the brane modes may be 
exponentially suppressed. 
However, some of them may sit on the branes. For $p>3$, some of
these modes may even be localized inside the branes,
{\em i.e.}, in the dimensions which are compactified.
In contrast to the KK modes (from either the branes or the bulk),
these localized modes can be produced as a resonance
from the scattering of two brane modes which have no momentum in
the compactified dimensions, with a significant
({\em i.e.}, not gravitational strength) coupling.
Some of these localized modes are the pseudo-Goldstone 
bosons eaten by the anomalous $U(1)$ gauge bosons 
mentioned above. Experimental constraints require the remaining ones to pick up
relatively large masses, but one of them may behave like an axion.

{}For example, in an orbifold string model, these fields are the twisted 
sector moduli which are stuck at the orbifold fixed points (or fixed lines). 
Unlike the untwisted sector
states which are free to propagate in the bulk, 
the twisted sector states have no momentum or winding in
the compactified dimensions, and hence are localized in the internal space.
While the couplings of the brane fields with the bulk fields 
are suppressed by a factor of 
$(V_{p-3} V_{9-p} M_s^6)^{-1}=(M_s/M_P)^2$ 
due to wavefunction normalization in the compactified
dimensions (and hence the couplings are of
gravitational strength), this suppression factor (or part of it) 
is absent in the corresponding coupling with these twisted fields.
If these fields happen to sit on the branes, the
coupling with the brane fields is simply $\kappa \sim g_s$ and is 
therefore independent of $M_s$, $V_{p-3}$ and $V_{9-p}$. 
Before SUSY breaking, these fields, being moduli, are massless.
However, to be compatible with experiment, these states must pick up masses.
Assuming that dynamical SUSY breaking takes place
on the branes we expect these twisted fields to obtain masses
somewhat comparable to the SUSY breaking scale, $M_{SUSY} \sim O(1)$ TeV. 

In the scattering of two brane modes with no momentum in the compactified
dimensions, the brane KK modes cannot be singly produced, since this would
violate momentum conservation along the branes. 
In contrast, the momenta
orthogonal to the branes need not be conserved since the branes break
translation invariance in these orthogonal directions. Therefore, the bulk KK
modes can be singly produced by the brane modes, with a coupling in the 
amplitude $\propto 1/M_P$.
After the summation over the large multiplicity $\sim (s^{1/2}R)^n$ of bulk KK
modes (where $n$ denotes the number of large compactification dimensions
transverse to the branes with size $\sim R$, and $s$ denotes the CM mass 
energy squared), the cross section for processes
involving real bulk KK emission is $\propto 1/M_{Pl}^{n+2}$.  For processes
involving the exchange of virtual bulk KK modes, summing over the exchanges
that do not have strong propagator suppression,
the amplitude is $\sim (1/M_P^2)(s^{1/2}R)^n
\propto 1/M_{Pl}^{n+2}$, and the resultant cross sections 
are $\propto 1/M_{Pl}^{2(n+2)}$.
Given that $M_{Pl} \gsim 30-50$ TeV ($n=2$), these effects may be too small to
be observable at Tevatron or even LHC energies.  In contrast to the KK modes,
the twisted modes which have no momentum or winding in the compactified
dimensions can be singly produced by the brane modes with a coupling
$\propto 1/M_s$. They can thus be produced as resonances with strengths that
are not suppressed by the size of the extra large
dimensions.  Despite the fact that the
couplings of these twisted fields with the Standard Model fields do not appear
at the renomalizable level, and the cross sections for processes involving the
exchange of these twisted fields are hence suppressed by the factor $1/M_s^4$,
their experimental signatures may well be larger than those of the gravity KK
modes.

In a typical Type I string model, 
there are abelian gauge fields (brane modes) with field-theoretic triangle
anomalies.  They become massive generically, with some of the above-mentioned 
twisted RR scalars being the would-be-Goldstone bosons.  
If the mass of an abelian gauge field happens to be in the LHC energy 
regime,
it may be seen as a resonance in the hadron scattering experiment 
and in the Drell-Yan process. 
Since these massive gauge bosons are
brane fields, they couple to the
quark-antiquark and lepton-antilepton pair with the strength comparable
to that of the SM interactions.
In contrast to perturbative heterotic string theory, 
there can be more than one anomalous $U(1)$ gauge fields for Type I strings. 
Moreover, these $U(1)$ gauge fields can be relatively light
and are possible candidates of the $Z^{\prime}$ bosons.

\section{Some Generic Features of the Brane World}

{}For illustrative purposes, let us first consider compactification of
Type I string theory to 4D on a ${\mathbb Z}_3$ orbifold \cite{z3}. 
The ${\mathbb Z}_3$
generator $g$ acts on the complex
coordinates $z_1$, $z_2$, $z_3$ of the compactified dimensions 
$T^6=T^2 \times T^2 \times T^2$ as follows:
$g z_i = \omega z_i$ for $i=1,2,3$ 
where $\omega=e^{2i \pi/3}$. The resulting model has ${\cal N}=1$
supersymmetry. The twisted sector states are located at points
in the compactified dimensions
that are invariant under the orbifold action.
Here there are $27$ fixed points at which the
${\mathbb Z}_3$ twisted sector states are located: 
$(z_1,z_2,z_3)=(1/\sqrt{3})e^{\pi i/6}( k_1 R_1,k_2 R_2 ,k_3 R_3)$, 
where $k_i=0,1,2$ and $R_i$ are the radii of the three $T^2$'s 
(in each $T^2$, the two radii are the same because of the ${\mathbb Z}_3$
orbifold symmetry).
The twisted sector states at each fixed point form
an ${\cal N}=1$ neutral chiral supermultiplet.
In particular, the scalar component is $\phi_k + i a_k$ where
$\phi_k$ ($a_k$) comes from the NS-NS (R-R) sector. 

Consistency ({\em i.e.},
tadpole cancellation) requires introducing open string and
$32$ $D9$-branes.
In the T-dual picture where
two of the complex dimensions $z_1$ and $z_2$ are dualized,
the $D9$-branes become $D5$-branes. 
The gauge group is maximal when
the $D5$-branes are sitting on top on each other, for instance, when they
are all located at $z_1=z_2=0$.
In this case, the model has $SO(8) \times SU(12) \times U(1)$ gauge
symmetry. The open string sector gives rise to
three families of charged chiral multiplets in the 
$({\bf 8},\overline{\bf 12})(-1)$ and $({\bf 1},{\bf 66})(+2)$
representations.
In this brane configuration, not all the twisted sector states 
are located on the branes.
In particular, out of the $27$ twisted sector chiral multiplets, 
only $3$ of them (the ones with $k_1=k_2=0$)
are located on the branes. The remaining twisted sector states 
are located at a non-zero distance from the branes.

{}The strength of the coupling of a twisted field with the brane
fields depends on the 
number of dimensions in which the twisted field lives, and its
location from the branes. 
In the above example, the ${\mathbb Z}_3$ twisted sector states 
live in four dimensions and so for the corresponding fixed points
which lie on the branes, the couplings with
the brane modes are unsuppressed, {\em i.e.}, $\kappa \sim g_s$.
For the remaining twisted moduli located
at non-zero distances
$|X|$ from the branes, the corresponding couplings with the brane fields
are exponentially suppressed as $\exp(-(M_s |X|)^2)$.

\begin{figure}
\centering
\epsfxsize=3 in
\hspace*{0in}\vspace*{.2in}
\epsffile{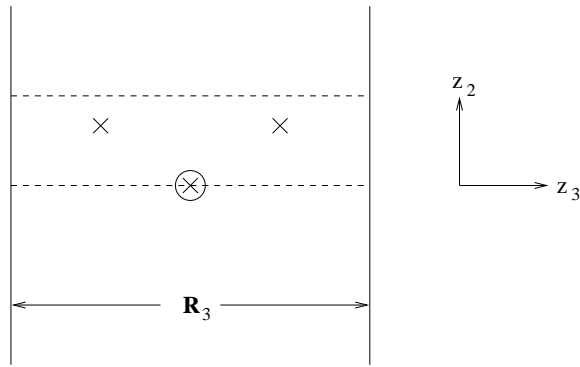}
\caption{\footnotesize{A schematic diagram of various types of twisted fields.
The complex coordinate $z_1$ is suppressed in this diagram.
The ${\mathbb Z}_3$ and ${\mathbb Z}_6$
twisted fields are localized at fixed points indicated by
crosses and a circle respectively. 
The ${\mathbb Z}_2$ twisted field are 
localized at fixed loci indicated
by dotted lines.
The two solid lines which are separated by $R_3$ (the size of the 
compactified dimensions in the $z_3$ direction) are identified.
Our brane world is the dotted line with $z_1=z_2=0$
and so the ${\mathbb Z}_6$ fixed point is localized on our brane.}}
\label{twistedfig}
\end{figure}

{}In more realistic string models, there can be more
than one type of twisted fields. In particular, there can be 
some twisted fields which are located
at fixed loci rather than fixed points
in the internal space. For example, in the ${\mathbb Z}_6$ orbifold \cite{z6}
generated by the above ${\mathbb Z}_3$ twist and an
additional ${\mathbb Z}_2$ twist where the 
${\mathbb Z}_2$ generator $R$ acts on the coordinates as follows:
$R z_1 = - z_1$ , $R z_2 = - z_2$ , and $R z_3= z_3$. 
The ${\mathbb Z}_2$ twisted sector states are located
at $16$ fixed loci in the internal space with $z_1$ and
$z_2$ given by
$(z_1,z_2)= {1 \over 2} \left( (a_1+ib_1)R_1, (a_2+ib_2)R_2 \right)$
where $a_i,b_i=0,1$.
While the ${\mathbb Z}_2$ twisted fields have no momentum or winding 
in the
complex dimensions $z_1$ and $z_2$, 
they are free to propagate in the complex dimension not twisted by the
${\mathbb Z}_2$ generator, {\em i.e.}, $z_3$.
Therefore, the ${\mathbb Z}_2$ twisted sector states
live in six dimensions. Even for the ${\mathbb Z}_2$ twisted
states that sit on the branes (the ones with $a_2=b_2=0$),
the corresponding couplings are suppressed
by the volume of the compactified dimensions in which the
${\mathbb Z}_2$ twisted sector states are free to propagate.
In other words, $\kappa \sim g_s / (V_3 M_s^2)$
where $V_3$ is the volume of the compactified dimensions not twisted by
${\mathbb Z}_2$.
In addition to the 
${\mathbb Z}_3$ and ${\mathbb Z}_2$ twisted fields described above, 
there are also
${\mathbb Z}_6$ twisted sector states.
A schematic diagram of various types of twisted fields
is given in Figure \ref{twistedfig}.

{}The gauge group of this model is 
$[SU(6)\otimes SU(6) \otimes SU(4) \otimes U(1)^3]^2$.
Although this contains the SM, the
residual gauge symmetry is too large for the model to be phenomenologically
interesting. The rank of the gauge group can be reduced by turning on the
untwisted NS-NS sector B-field background \cite{bij}. In particular,
a three-family Pati-Salam like model was constructed in this 
framework \cite{ST}. Some other four-dimensional ${\cal N}=1$ 
Type I string models were presented in Ref. \cite{zk}.
As noted above, the twisted fields are massless before SUSY is
broken, but would pick up masses
after SUSY breaking and dilaton stabilization.

The coupling of the twisted R-R fields $a_k$ to a $U(1)$ gauge field is 
proportional to ${\mbox{Tr}} (\gamma_k \lambda_i) ~\partial_{\mu} a_k 
A_i^{\mu}$, 
where $\lambda_i$ is the Chan-Paton wavefunction of the abelian gauge field
$A_i^{\mu}$, 
and $\gamma_k$ defines the action of the
orbifold on the D-branes.
The coupling of $a_k$ with a pair of
gauge fields (abelian or non-abelian) is $\propto 
{\mbox Tr} (\gamma_k^{-1} \lambda_G^2) ~ a_k F \tilde{F}$ 
where $\lambda_G$ is the Chan-Paton wavefunction of the gauge fields
with gauge group $G$. Here, $F$  ($\tilde F$) is the
field strength (dual field strength) for the abelian or non-abelian gauge
fields.  If $G$ is strongly coupled, the twisted
R-R fields $a_k$ can play the role of axions.
In contrast to perturbative heterotic string theory, a generic Type I 
string model can have more than one anomalous $U(1)$ \cite{ibanez}. 
Their triangle anomalies are cancelled by the generalized Green-Schwarz
mechanism \cite{gs} which involves the exchange of the R-R axions.
It is well known that the coupling which mixes a R-R axion with a $U(1)$ gauge
field is part of the $U(1)$ gauge-invariant combination 
${1\over 2} {M_s^2} (\partial_{\mu} a_i - g_i e_i A_i^{\mu})^2$
where $A_i^{\mu}$, $i=1,2,...N$ are the anomalous $U(1)$'s, $g_i$
are the gauge couplings and $e_i$ 
are the corresponding charges.
The $N$ $U(1)$ gauge bosons become massive, eating up $N$ of the 
twisted R-R fields (more precisely, $N$ linear combinations
of the twisted R-R fields $a_k$ described above). 

Of the remaining twisted RR fields, it is likely that all except 
one are relatively heavy. Although we do not expect continuous global 
symmetry in a string model, it is plausible that some approximate global 
symmetries are present in the low energy effective field theory, 
{\em i.e.}, such symmetries are broken only by high-dimensional operators in 
the effective theory. If an approximate Peccei-Quinn symmetry is 
present, effectively, we have an axion field. 
In this situation, strong interaction chiral symmetry dictates its 
properties, that is, its mass and coupling follow from its mixing with the 
neutral pion, with the mixing given by $f_{\pi}/M_s$. For $M_s \sim 10$ TeV, 
this is clearly ruled out by experiments\cite{kim}. However, typically, 
this axion field will also couple to other non-abelian gauge fields in 
the hidden sector, and their gauge couplings are expected to be strong 
at relatively high scales (compared to the QCD scale). In this situation,
the axion will pick up a relatively large mass and so can avoid 
detection \cite{tye}.

{}Let us now focus on the couplings of the twisted states with the SM
fields. The bosonic states from the twisted sectors have spin-$0$, and
so can couple to a pair of gauge fields. In particular, the above mentioned
couplings 
of the R-R axions with a pair of gauge fields  
belong to this type. Similarly, the 
spin-$0$ fields from the NS-NS sectors also couple to the gauge fields,
in the form 
\begin{equation}
{\mbox Tr} (\gamma_k^{-1} \lambda_G^2) ~ \phi_k F^2 ~. 
\end{equation}

{}The spin-$0$ fields from the twisted sectors
can also couple to 
a pair of fermions ${\psi}_L$ and $\chi_R$ with opposite chirality. 
Since the twisted fields (denoted collectively as $T$)
are gauge singlets, the three-point coupling to the Standard Model fermions 
$T \overline{\psi}_L \chi_R$ is absent, since it is not a Standard Model 
singlet.
However, the lowest order non-renormalizable couplings, such as
\begin{equation}
{\cal C}_{Qd} {g_s^{\footnotesize{3/2}} \over M_s} \overline{Q}_L d_R H_d T, \ \ 
{\cal C}_{Qu} {g_s^{\footnotesize{3/2}} \over M_s} \overline{Q}_L u_R H_u T, \ \ 
{\cal C}_{L\ell} {g_s^{\footnotesize{3/2}} \over M_s} \overline{L}_L \ell_R H_d T, \ \ 
\label{ffht}
\eeq
are presumably allowed \footnote{The powers of $g_s$ can be seen as follows.
The coupling between three open string states is proportional to
$g_s^{1/2}$ whereas the coupling of one closed and two open string states
is proportional to $g_s$. The above
$4$-point terms can be factorized into a product of these two
types of three-point couplings.},
are presumably allowed, where, in standard notation, $Q={u \choose d}$, 
$L= {\nu_\ell \choose \ell}$, $H_{d,u}$ are the $Y=1,-1$ Higgs in the
MSSM, and we suppress generation labels 
(the ${\cal C}$'s are dimensionless coupling matrices in generation space). 
There are two types of cubic couplings that result from these quartic
couplings.  First, nonperturbative effects can generate a nonzero 
potential for the $T$ fields and hence they can acquire non-zero vevs.
These produce the cubic couplings
\begin{equation}
 {\cal C}_{Qd} 
   {g_s^{\footnotesize{3/2}} \over M_s} \overline{Q}_L d_R H_d \langle T \rangle, \ \ 
 {\cal C}_{Qu} 
   {g_s^{\footnotesize{3/2}} \over M_s} \overline{Q}_L u_R H_u \langle T \rangle, \ \ 
 {\cal C}_{L\ell}
   {g_s^{\footnotesize{3/2}} \over M_s} \overline{L}_L \ell_R H_d \langle T
\rangle 
\label{yukawa}
\end{equation}
These would combine with the original dimension-$4$ Yukawa operators, and the
generational/flavor structure of the matrices ${\cal C}$ may help to yield
the observed hierarchical fermion masses and quark mixing. 
Second, when $H_u$ and $H_d$ pick up vevs,
the quartic operators in (\ref{ffht}) will yield the cubic operators
\begin{equation}
 {\cal C}_{Qd} 
    {g_s^{\footnotesize{3/2}} \over M_s} \overline{Q}_L d_R \langle H_d \rangle T, \ \
 {\cal C}_{Qu} 
    {g_s^{\footnotesize{3/2}} \over M_s} \overline{Q}_L u_R \langle H_u \rangle T, \ \
 {\cal C}_{L\ell}
    {g_s^{\footnotesize{3/2}} \over M_s} \overline{L}_L \ell_R \langle H_d
\rangle T
\label{3point}
\end{equation}
For simplicity, let us neglect mixing effects. If 
the ${\cal C}$ have the structure of the quark and lepton mass hierarchies, 
$\sim m_f/M_{ew} << 1$ for all fermions $f$ other than the top quark, then the
induced 3-point couplings in (\ref{3point}) are presumably rather small. 
Note that in the Pati-Salam case in Ref \cite{ST}, the right-handed fermion
components $f_R$ are $SU(2)_R$ doublets, which generically gives rise to a 
further suppression factor $\propto M_R/M_s$, where $M_R$ denotes the 
mass of the $W_R$ vector boson.  

Thus, some linear combinations of the twisted RR scalars are eaten by
the anomalous $U(1)$ gauge bosons. The remaining ones $a_i$ together with
the twisted NS-NS scalars $\phi_i$ can appear in the low
energy effective action. Some of the 
lowest order terms in the effective Lagrangian involving
the couplings of these twisted modes or the anomalous $U(1)$
gauge fields with the SM fields are:
\begin{eqnarray}\label{leff}
{\cal L} =&& 
       {g_s \over M_s} \left(
            \sum_{i,j} {\cal C}^{(1)}_{ij} \phi_i F_j^2 
           + \sum_{i,j} {\cal C}^{(2)}_{ij} a_i F_j \tilde{F}_j \right) 
      +   {g_s^{3/2} \over M_s} \left(
          \overline{f}_L f_R H \sum_{i,f} {\cal C}^{(3)}_i a_i
      +   \overline{f}_L f_R H \sum_{i,f} {\cal C}^{(4)}_i \phi_i
          \right) \nonumber \\
   + &&  \left( 
          \sum_{i,f} g_i e_{Li} \overline{f}_L \gamma_{\mu} f_L A^{\mu}_i 
        + \sum_{i,f} g_i e_{Ri} \overline{f}_R \gamma_{\mu} f_R A^{\mu}_i
        \right) + \dots
\end{eqnarray}
where $\overline{f}_L f_R H=\overline{Q}_Ld_RH_d$, $\overline{Q}_Lu_RH_u$, 
$\overline{L}_L\ell_RH_d$, generation
indices are suppressed, ${\cal C}^{(k)}_{ij}$ and ${\cal C}^{(k)}_i$ are 
model-dependent coefficients, 
$A_i^{\mu}$ denote the anomalous $U(1)$ gauge fields, with gauge 
couplings $g_i$, and $e^L_i$ and $e_i^R$ are the corresponding charges. 
$F_j^2$ and $F_j \tilde F_j$ sum over abelian and
nonabelian gauge fields.  
The nonperturbative effects that generate a nonzero potential for the moduli
(as is necessary for dilaton stablization and is related to SUSY breaking)
will, in general, produce various quadratic, cubic, and quartic couplings
involving these moduli.  These are subject to obvious constraints; for example,
they must be invariant under the orbifold twists.  

In this letter, we discuss Type I strings on orbifolds, and
more generally, 
when the orbifold singularities are smoothed out (``blown-up'') to a Calabi-Yau
manifold as the twisted NS-NS scalars $\phi_k$ acquire non-zero vevs. The
coupling of the twisted modes with the brane fields is 
$\kappa \sim g_s f(\langle \phi_k \rangle)$ where $f(\langle \phi_k \rangle)$
is a function of $\langle \phi_k \rangle$. 
Even in the generic case where $\langle \phi_k \rangle$ is comparable with the
string scale, it is likely that the suppression factor for the bulk fields,
{\em i.e.}, $(V_{p-3} V_{9-p} M_s^6)^{-1}$ is still much smaller than
$f(\langle \phi_k \rangle)$ and so there are closed string fields which couple
to the brane fields more strongly than the bulk fields.  Moreover, we expect
that the features we discuss here would still apply for other realizations of
the brane world (which may be viewed as dual to the Type I description), such
as non-perturbative heterotic string/M theory models with solitonic fivebranes
\cite{ovrut} as well as F-theory \cite{vafa} models.

\section{Experimental Signatures and Constraints}

There are several experimental implications of eq. (\ref{leff}). 
In general, there could be couplings such as $g_s\mu_{ijk} T_iT_jT_k$, where 
the coefficient $\mu_{ijk}$ would presumably be of order the electroweak scale.
These would allow a subset of the $\xi = \phi_k,a_k$ to decay rapidly, with
widths $\Gamma \sim g_s^2\mu^2/m_\xi$.  We concentrate here on the light $\xi$
fields for which these decay channels are kinematically forbidden, as
well as the subset of the heavier $\xi$ fields whose analogous decays do not
occur because the corresponding cubic or quartic operators are absent 
(as a consequence, for example, of noninvariance under the orbifold twists). 

In the following, we refer collectively to the various
mass eigenstates of $\phi$ and $a$ and suppress their multiplicity; similarly,
we use ${\cal C}$ to refer to the appropriate linear combinations of
the original ${\cal C}_{ij}^{(1)}$ and ${\cal C}_{ij}^{(2)}$ for these 
mass eigenstates. 
The masses
$m_\xi$, $\xi=\phi,a$, are expected to be of roughly of order the SUSY breaking
scale, which is comparable to the electroweak symmetry breaking scale.  The
couplings in (\ref{leff}) give rise to the decays $\xi \to gg$ as well as 
$\xi \to \gamma\gamma$, and (if kinematically allowed) $\xi 
\to ZZ, \ W^+W^-$.  From the lowest-order graphs, in terms of 
\beq
\Gamma_0 = \frac{g_s^2 {\cal C}^2}{64 \pi} \Bigl 
(\frac{m_\xi}{M_s}\Bigr )^2 m_\xi
\label{gammat}
\eeq
we find $\Gamma(\xi \to gg)= (N_c^2-1)\Gamma_0$ where $N_c=3$, 
$\Gamma(\xi \to \gamma\gamma)=\Gamma_0$, 
 $\Gamma(\xi\to ZZ)=\Gamma_0(1-4m_Z^2/m_\xi^2)^{1/2}$, and
 $\Gamma(\xi \to W^+W^-)=2\Gamma_0(1-4m_W^2/m_\xi^2)^{1/2}$, 
with $\xi=\phi,a$. 
Since one expects $g_s \sim O(1)$, these lowest-order calculations are only 
rough estimates.  Summing the partial widths and taking $g_s \sim O(1)$, 
${\cal C} \sim O(1)$, and $m_{W,Z}^2/m_\xi^2 <<1$, we get 
\beq
\Gamma_\xi \sim 0.1(m_\xi/M_s)^2m_\xi
\label{gammatotphi}
\eeq
For $m_\xi=1$ TeV, $M_s=10$ TeV, this gives $\Gamma_\xi \sim
10^{-3}m_\xi \sim 1$ GeV.  The dominant decays of the $\phi,a$, namely
$(\phi,a) \to gg$, should not involve large missing energy and should in
principle allow one to reconstruct the mass.  

The couplings $\phi F^2$ and $a F \tilde F$ with the gluon field 
in (\ref{leff}) contribute to single jet inclusive, dijet, and multijet
production in $\bar p p$ (and, equally, in $pp$) collisions.  The $\xi$
can be produced in the process $gg \to g \xi$ involving the usual triple
Yang-Mills vertex multiplied by $g_s{\cal C}/M_s$.  The resultant decay of the
$\xi$ yields 3-jet events.  The rate, relative to regular QCD 3-jet events is
roughly $(g_s{\cal C}/\alpha_3)^2({\hat s}/M_s)^2(1-m_\xi^2/{\hat s})^{1/2}$
(where $\hat s$ is the $gg$ center of mass energy squared) which could be 
$O(10^{-2})$ for $\sqrt{s}=1.8$ TeV and our illustrative values 
of $g_s, M_s, m_\xi$.  

In $gg \to gg$ scattering, there are new contributions from graphs involving 
the coupling of $gg$ to $\xi=\phi,a$ in the $s,t$ and $u$ channels.  From the
tree-level graphs, we compute the new contribution to $gg \to gg$ to be 
$d\sigma/d\hat t = |{\cal M}|^2/(16 \pi \hat s^2)$ where 
\beqs
& & \overline{\sum}|{\cal M}|^2 = 
\frac{(N_c^2-1)}{16}\Bigl (\frac{g_s{\cal C}}{M_s} \Bigr )^4
\sum_{m=m_\phi,m_a} \biggl [
\frac{\hat s^4}{(\hat s-m^2)^2} + 
\frac{\hat t^4}{(\hat t-m^2)^2} + 
\frac{\hat u^4}{(\hat u-m^2)^2} + \cr\cr
& & \frac{\hat s^4+\hat t^4+\hat u^4-2\hat u^2(\hat s^2+\hat t^2)}
{2(\hat s-m^2)(\hat t-m^2)} + 
\frac{\hat s^4+\hat t^4+\hat u^4-2\hat s^2(\hat t^2+\hat u^2)}
{2(\hat t-m^2)(\hat u-m^2)} +
      \frac{\hat s^4+\hat t^4+\hat u^4-2\hat t^2(\hat s^2+\hat u^2)}
{2(\hat s-m^2)(\hat u-m^2)} \biggr ]
\label{gg}
\eeqs
where integration over final state phase space involves a (1/2) factor for
identical particles. (As indicated, the $\phi F^2$ and $a F \tilde F$
contributions are of the same form, with $m_\phi \to m_a$.)  
Since $g_s \sim O(1)$, higher order corrections can be substantial; however,
(\ref{gg}) gives a rough estimate.  For $\hat s 
\simeq m^2$, the $s$-channel term has a strong 
resonant enhancement (since the $s$-channel propagator is actually 
$\propto 1/(m \Gamma)$ at $\hat s=m^2$), 
and, in particular, the $M_s^{-4}$ factor from the
coupling is cancelled by the $M_s^4$ factor in $\Gamma_\xi^{-2}$ arising from
the square of the $s$-channel propagator.  Inserting (\ref{gammatotphi}), we
find that $\overline{\sum}|{\cal M}|^2 \simeq 50(g_s{\cal C})^4$.  This is
comparable to the regular QCD contribution, which, e.g., for $gg$ CM angle
$\hat \theta=\pi/2$, gives $\overline{\sum}|{\cal M}|^2 \simeq 30 g_3^4$, 
where $g_3$ is the SU(3) coupling. The most striking effect is that there
would be peaks in the dijet invariant mass at $M_{JJ}=m_{\phi,a}$.  
We recall
that there are, in general, several $\phi_k$ and $a_k$, so there could be
several such resonances.  
Let us compare the size of this effect with 
the conventional scenario in which $M_s \sim M_P$.
In that case, the moduli fields $\phi,a$ could also be of order the SUSY
breaking scale, and, given that $M_s$ cancels out at the resonance for 
$\hat s = m_\xi^2$, the existence of this resonance, {\it per se}, would not 
constitute evidence of low-string scale, as opposed to conventional $M_s \sim
M_P$ models; however, the contribution of (\ref{gg}) to 
$d^2\sigma/dM_{JJ}d\cos \hat \theta$ involves the convolution
over the momentum fractions $x_1,x_2$ of the gluons, and the off-resonant 
contributions to this convolution would be very different for low- and 
high-string scale theories because of the quite different $M_s$ scales. 

Current data on 2-jet and inclusive jet production in $\bar p p$ 
collisions from D0 is in excellent agreement with QCD predictions
\cite{d0jets}.  This is also true of the CDF data in the region where the $gg
\to gg$ subprocess makes its main contribution, for $E_T \lsim 300$ GeV 
\cite{jetrev} (the latter resulting from the rapid rise of the gluon 
distribution functions $g(x)$ for small $x$ and the fact that $E_T^2 = (\hat
s/4)\sin^2 \hat \theta < x_1 x_2 s/4$). Although the new contributions from
$\phi,a$ exchange can only roughly be estimated, in view of the expected large
$g_s \sim O(1)$ and the model-dependent factor ${\cal C}$, we infer that if
${\cal C} \sim 1$, then a safe bound in order to avoid 
conflict with this data is
$m_{\phi,a} \gsim O(1)$ TeV.  Furthermore, if $m_{\phi,a}$ are sufficiently
large so that there is no significant resonant contribution, the data would
still constrain $M_s \gsim O(1)$ TeV, given that the corresponding couplings
$g_s {\cal C} \sim O(1)$.  In models with a low string scale and large
compactification radii, there will also be a new contributions to $\bar p p$
scattering from processes involving the exchange or emission of gravitational
KK modes; however, in a global data analysis, one could still distinguish
between the effects of these KK modes and of the $\phi,a$.  

Besides the contribution of the $\phi_k F_j^2$ term to gauge couplings, 
via $\langle \phi_k \rangle$, 
these terms also contribute to various loop effects which could
be significant.  
Another important implication concerns neutral vector bosons. 
In contrast to perturbative heterotic string theory, 
there can be several anomalous $U(1)$ gauge symmetries in Type I string
models.
Moreover, in perturbative heterotic string theory where
$M_s \sim M_P$, the associated anomalous $U(1)$ vector bosons are
too heavy to be detected.
However, the 
anomalous $U(1)$ vector bosons in the present scenario can be relatively
light ($\lsim$ few TeV), and are possible candidates for
$Z^{\prime}$ bosons.  These vector bosons gain masses by ``eating'' some of the
above axions.  The resultant $Z^\prime$'s couple to the brane fields with
strength comparable with SM interactions.  Current lower bounds on such 
$Z^{\prime}$ bosons are of order 800 GeV, depending on their couplings 
\cite{rpp}.  

In addition to the scalar component fields $\phi,a$, the twisted moduli chiral
superfield ${\cal T}$ also has modulino component fields $\tilde \phi,
\tilde a$. We denote these collectively as $\tilde T$. Clearly these would 
play the role of a electroweak singlet (``sterile'') neutrinos.  Once a nonflat
superpotential is generated (nonperturbatively) for the twisted moduli, there
would be terms of the form ${\cal T}{\cal T}$ yielding the bilinears $\tilde
T^T C \tilde T$, with mass coefficients presumably of order the SUSY breaking
scale and hence comparable to the electroweak scale.  If allowed by the matter
parities resulting from the underlying string theory, there could also be cubic
${\cal L}{\cal T}{\cal H}$ terms in the superpotential, which would give rise
to Dirac neutrino mass terms of the form $\bar \nu_L \tilde T_R$.  Note that
the $\tilde T$ fields couple to brane fields, in particular,
neutrinos, without the volume suppression factors affecting the couplings of
the bulk fields to brane fields.  We shall discuss this further elsewhere. 
This contrasts with the situation for
candidates for sterile neutrinos coming from the bulk \cite{neutrino}. 

The role of a generation dependent 
anomalous $U(1)$ gauge symmetry in the
observed hierarchical fermion mass structure has been studied in the
context of perturbative heterotic string theory \cite{u1a}.
An important difference here is that 
there can be
several anomalous $U(1)$'s (for example, the $Z_6$ model discussed
earlier has $4$ anomalous $U(1)$'s), and different families have different
$U(1)$ charge assignments, as in the case in Ref \cite{ST}.
Hence the possibilities in the brane world are more
intricate.

The numerous testable implications of models with low
string scale are clearly of great interest and
deserve further theoretical and experimental investigation.

\acknowledgments

{}We thank Philip Argyres, Zurab Kakushadze and Piljin Yi for 
discussions. The research of G.S. and R.S. is partially supported by the NSF
grant PHY-97-22101. The research of S.-H.H.T. is partially supported by
the NSF.

\end{document}